\documentclass[pre,aps,amsfonts,amsmath,superscriptaddress,showpacs,floatfix,twocolumn]{revtex4}
\usepackage{graphicx,epsfig}
\usepackage{ulem}
\usepackage{color}

\newcommand{\rr}{{\bf r}}

\newcommand{\cZ}{{\mathcal Z}}

\begin{document}

\title{Equivalence between fractional exclusion statistics and self-consistent mean-field theory in interacting particle systems in any number of dimensions}

\author{D. V. Anghel}
\affiliation{Horia Hulubei National Institute for Physics and Nuclear Engineering, 077126 M\u agurele, Ilfov, Romania}

\author{G. A. Nemnes}
\affiliation{Horia Hulubei National Institute for Physics and Nuclear Engineering, 077126 M\u agurele, Ilfov, Romania}
\affiliation{University of Bucharest, Faculty of Physics, ``Materials and Devices for Electronics and Optoelectronics'' Research Center, 077125 M\u agurele-Ilfov, Romania}

\author{F. Gulminelli}
\affiliation{LPC/ENSICAEN, 6 Bd du Mar\'echal Juin, 14050 Caen Cedex, France}

\begin{abstract}
We describe a mean field interacting particle system in any number of dimensions and in a generic external potential as an ideal gas with fractional exclusion statistics (FES).
We define the FES quasiparticle energies, we calculate the FES parameters of the system and we deduce the equations for the equilibrium particle populations. 
The FES gas is ``ideal,'' in the sense that the quasiparticle energies do not depend on the other quasiparticle levels populations and the sum of the quasiparticle energies is equal to the total energy of the system. We prove that the FES formalism is equivalent to the semi-classical or Thomas Fermi limit of the self-consistent mean-field theory and the FES quasiparticle populations may be calculated 
from the  Landau quasiparticle populations by making the correspondence between the FES and the  Landau quasiparticle energies. 
The FES provides a natural semi-classical ideal gas description of the interacting particle gas.
\end{abstract}


\maketitle

\section{Introduction} \label{intro}
Quantum (Bose or Fermi) statistics is a very basic concept in physics, directly arising from the indiscernability principle of the microscopic world. However effective models employing different statistics have proved to be
very useful to describe some selected aspects of complex interacting microscopic systems. 
As an example, an ideal Fermi gas with Fermi-Dirac statistics  can be described in certain approximations by a Boltzmann distribution with repulsive interaction, while a bosonic ideal gas, by an attractive potential \cite{NuclPhysB92.445.1975.Chaichian}.
On more general grounds,  the statistical mechanics of  fractional exclusion statistics (FES) quasi-particle systems was formulated by several authors
\cite{PhysRevLett.72.600.1994.Veigy,PhysRevLett.73.922.1994.Wu,PhysRevLett.73.2150.1994.Isakov,FES_intro2013.Murthy}, based on the FES concept
 introduced by Haldane in Ref. \cite{PhysRevLett.67.937.1991.Haldane}.
%
This has been a prolific concept and was applied to both quantum and classical systems (see e.g. \cite{PhysRevLett.67.937.1991.Haldane,PhysRevLett.73.922.1994.Wu,PhysRevLett.73.2150.1994.Isakov,
NewDevIntSys.1995.Bernard,PhysRevE.75.61120.2007.Potter,
PhysRevE.76.61112.2007.Potter,PhysRevE.84.021136.2011.Liu,
PhysRevE.85.011144.2012.Liu,JStatMech.P04018.2013.Gundlach,
PhysRevB.56.4422.1997.Sutherland,PhysRevLett.85.2781.2000.Iguchi,
JPhysA.40.F1013.2007.Anghel,PhysLettA.372.5745.2008.Anghel,PhysLettA.376.892.2012.Anghel,
EPL.90.10006.2010.Anghel,PhysRevLett.74.3912.1995.Sen,
JPhysB33.3895.2000.Bhaduri,PhysRevLett.86.2930.2001.Hansson,
PhysRevE.82.031137.2010.Mirza,PhysRevE.83.021111.2011.Qin,PhysRevE.76.061123.2007.Pellegrino,JPhysA.26.4017.1993.Chaichian}).

A stochastic method for the simulation of the time evolution of FES systems was introduced in Ref. \cite{JStatMech.P09011.2010.Nemnes} as a generalization of a similar method used for Bose and Fermi systems \cite{JStatMech.2009.P02021.2009.Guastella}, whereas the relatively recent experimental realization of the Fermi degeneracy in cold atomic gases has renewed the interest in the theoretical investigation of non-ideal Fermi systems at low temperatures and their interpretation as ideal FES systems
\cite{JPhysB42.235302.2009.Bhaduri,JPhysB.43.055302.2010.Qin,PhysRevE.83.021111.2011.Qin,JPhysA.45.315302.2012.vanZyl,JPhysA.46.045001.2013.MacDonald}.

The general FES formalism was amended to include the change of the FES parameters at the change of the particle species \cite{JPhysA.40.F1013.2007.Anghel,EPL.87.60009.2009.Anghel,PhysRevLett.104.198901.2010.Anghel,
PhysRevLett.104.198902.2010.Wu}. This amendment allows the general implementation of FES as a method for the description of interacting particle systems as ideal (quasi)particle gases \cite{PhysLettA.372.5745.2008.Anghel,PhysLettA.376.892.2012.Anghel,JPhysConfSer.410.012121.2013.Anghel,
PhysScr.2012.014079.2012.Anghel}.
However, the level of approximation of such a description and the connections with other many-particle methods are not yet clear. 

A rigorous connection between the FES and the Bethe ansatz equations
for an exactly solvable model was worked out in Refs.~\cite{PhysRevLett.73.3331.1994.Murthy,NewDevIntSys.1995.Bernard,PhysRevB.56.4422.1997.Sutherland}
in the one-dimensional case. In higher dimensions, one can expect that 
the statistics and interaction would not be transmutable in general because interaction and
exchange positions between two particles can occur separately. However indications exist 
that a mapping might exist between ideal FES and interacting particles with regular quantum statistics
in the quasi-particle semi-classical approximation given by the self-consistent mean-field theory. Indeed, in Refs. \cite{JPhysB33.3895.2000.Bhaduri,PhysRevLett.86.2930.2001.Hansson} the FES was applied to describe Bose gases with local ($\delta$-function) interaction in two-dimensional traps in the Thomas Fermi (TF) limit. In this paper we generalize this result to apply FES to Bose and Fermi
systems of particles with generic two-body interactions in arbitrary
external potentials in the Landau or TF semiclassical limit in any number
of dimensions.
We define quasiparticle energies which determine our FES parameters and using these we calculate the FES equilibrium particle distribution. Moreover, we calculate the particle distribution also starting from the mean-field description by defining the Landau type of quasiparticles, and we show that the two descriptions are equivalent, i.e. the populations are identical, provided that we make the mapping between the FES and Landau's quasiparticle energies. 
This equivalence proves that our FES description of the interacting particle system corresponds to 
a self-consistent mean-field approximation. Such an approximation, though certainly not adequate to fully describe strongly interacting quantum systems, is the basis of a number of highly predictive theoretical methods in many-body physics, from Landau Fermi liquid theory to density functional methods in correlated electron or nucleon systems. 

The structure of the paper is as follows.
In Section \ref{model} we introduce our model and calculate Landau's equilibrium particle population in the TF approach. In Section \ref{FES_section} we implement the FES description by using alternative quasiparticle energies and a definition of species. These species are related by the FES parameters, which we calculate. Using the FES parameters and quasiparticle energies we write the equations for the equilibrium particle populations. By making the correspondence between Landau's and the FES quasiparticle energies, we show that the FES equations are satisfied by Landau's populations. This proves that the FES formalism is consistent and suitable for the description of such interacting particle systems and that the FES quasiparticle populations may be calculated by Landau's approach, using the correspondence between the quasiparticle energies. In Sections \ref{num_ex} and \ref{O_ex} we show some numerical and analytical examples, respectively, whereas in Section \ref{Conclusions} we give the 
conclusions.

\section{The model in Landau's approach} \label{model}

Let us consider a system of $N$ interacting particles described by the generic Hamiltonian:
\begin{equation}
\hat{H}= \sum_{ij}(t_{ij}+V_{ij})\hat{a}_i^\dag\hat{a}_j+\frac{1}{2}\sum_{ijkl}v_{ijkl}\hat{a}_i^\dag\hat{a}_j^\dag\hat{a}_l\hat{a}_k \label{ham}
\end{equation} 
where the indexes $i,j,k,l$ denote the single particle states, and $\hat{a}^\dag_i (\hat{a}_i)$ are creation (annihilation) operators obeying commutations rules which define the quantum statistics of the system (Bose or Fermi). 
We assume that the particle-particle interaction $v(|\rr-\rr'|)$  depends only on the distance between the particles, whereas the external potential is $V_{\rm ext}(\rr)$. 
In the mean-field approximation the total energy of the system can be written as
\begin{equation}
  E = \sum_i (t_i+V_i) n^{(\pm)}_i + \frac{1}{2}\sum_{ij}v_{ij} n^{(\pm)}_in^{(\pm)}_j , \label{E_FLT}
\end{equation}
where $t_i=\langle i |\hat t |i\rangle$ are single-particle kinetic energies, ${V}_{i}=\langle i| \hat V_{\rm ext} |i\rangle$, and ${v}_{ij}=\langle ij| \hat v|ij\rangle \mp \langle ij| \hat{v}|ji\rangle$ are antisymmetrized (symmetrized) matrix elements for fermions (bosons). The upper and lower signs in the superscripts of the occupation numbers $n^{(\pm)}_i$ stand for fermions and bosons, respectively.

At the thermodynamic limit, the finite temperature properties of the system can be accessed via the grandcanonical partition sum defined by the mean-field one-body entropy as
\begin{eqnarray}
  \ln(\cZ)_{\beta\mu} &=& \mp\sum_i \{[1\mp n^{(\pm)}_i] \ln[1\mp n^{(\pm)}_i] \pm n^{(\pm)}_i \ln n^{(\pm)}_i\} \nonumber \\
  && - \beta (E-\mu N) , \label{cZ_def}
\end{eqnarray}
Maximizing this function with respect to the single particle occupations gives the equilibrium particle populations,
\begin{equation}
  n^{(\pm)}_i = \left[e^{\beta(\tilde{\epsilon}_i-\mu)}\pm 1\right]^{-1} , \label{npm_eq_L}
\end{equation}
where the quantities $\tilde{ \epsilon}_i \equiv \partial E/\partial n_i = t_i + V_i+\sum_j V_{ij} n^{(\pm)}_i$ are Landau's quasiparticle energies.

If  we assume a large number of particles, and a sufficiently slowly varying external potential, we can employ the Thomas-Fermi (or Landau) theory which amounts to extending this mean-field formalism to a finite inhomogeneous system employing a semi-classical limit for the kinetic energy. We can divide the system into macroscopic cells, $\delta\rr$, centered at $\rr$, where the external field is locally constant and apply the thermodynamic limit in each cell.
The single-particle energies are continuous variables and it is meaningful to introduce the density of states (DOS) in each cell as $\delta\rr\sigma(\rr,t)$. Then the quasi-particle energies read
\begin{eqnarray}
  \tilde\epsilon_{\rr}(t) &=& t +V_{\rm ext}(\rr)+ \int_\Omega d^s\rr' \int_0^\infty  v(|\rr-\rr'|) n^{(\pm)}(\rr',t') \nonumber \\
  && \times \sigma(\rr',t') \, dt' . \label{til_epsL}
\end{eqnarray}
We observe that the sum of the quasiparticle energies 
\begin{equation}
\tilde E=  \int_\Omega d^s\rr \int_0^\infty  \tilde\epsilon_{\rr}(t)  n^{(\pm)}(\rr,t)\sigma(\rr,t) \, dt
\end{equation}
is not equal to the total mean-field energy $E$, because of the well-known double counting of the interactions. 

\section{The FES formalism} \label{FES_section}

\begin{figure}[t]
\begin{center}
\includegraphics[width=7.5cm]{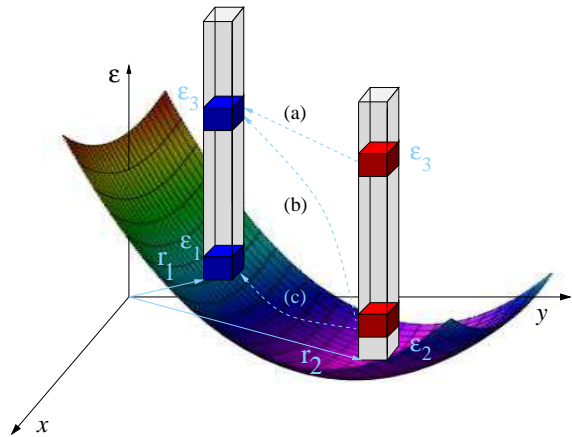}
\end{center}
\caption{
Species in a nonhomogeneous system in an external field, 
illustrating the change in the number of states in species located 
at $\rr_1$, upon inserting particles in species located at $\rr_2$. 
The FES parameters corresponding to Eqs. (14) are indicated: 
(a) $\alpha_{\delta\rr_1 \epsilon_3;\delta{\rr_2}\epsilon_3}$, 
(b) $\alpha_{\delta\rr_1\epsilon_3;\delta{\rr_2}\epsilon_2}$, and
(c) $\alpha_{\delta\rr_1\epsilon_1;\delta{\rr_2}\epsilon_2}$.
}
\label{species}
\end{figure}

Now we want to show that the semi-classical Thomas-Fermi results can be reproduced by an ideal system, provided that such a system obeys FES. A FES system consists of a countable number of species, denoted here by an index, $I$ or $J$ (we use capital letters to denote species). Each species $I$ contains $G_I$ available single-particle states and $N_I$ particles, each of them of energy $\epsilon_I$ and chemical potential $\mu_I$. If the particles are bosons, $G_I$ represents also the number of states in the species. If the particles are fermions, the number of single-particle states in the species $I$ is $T_I=G_I+N_I$. The FES character of the system consists of the fact that if the number of particles in one species, for example, in species $I$, changes by $\delta N_I$, then the number of states in any other species $J$ changes as $\delta G_J = \alpha_{JI}\delta N_I$ for bosons, or $\delta T_I = \alpha_{JI}\delta N_I$ for fermions. 
The parameters $\alpha_{IJ}$ are called the FES parameters. The total number of micro-configurations in the system is
\begin{subequations} \label{eqs_Ws}
\begin{equation}
  W^{(-)}_{\{(G_I,N_I)\}} = \prod_I \frac{(G_I+N_I-1)!}{N_I!(G_I-1)!} \label{eq_WB}
\end{equation}
for bosons and
\begin{equation}
  W^{(+)}_{\{(T_I,N_I)\}} = \prod_I \frac{T_I!}{N_I!(T_I-N_I)!} \equiv W^{(-)}_{\{(T_I-N_I+1,N_I)\}} \label{eq_WF}
\end{equation}
\end{subequations}
for fermions; the products in Eqs. (\ref{eqs_Ws}) are taken over all the species in the system. Using Eqs. (\ref{eqs_Ws}) and the fact that all the particles in a species have the same energy and chemical potential, we write the partition functions,
\begin{equation}
  \cZ^{(+)} = \sum_{\{N_I\}}\cZ^{(+)}_{\{(T_I,N_I)\}} \ {\rm and}\ \cZ^{(-)} = \sum_{\{N_I\}}\cZ^{(-)}_{\{(G_I,N_I)\}} \label{cZ_gen}
\end{equation}
where we used the notations
\begin{subequations} \label{cZBF_part}
\begin{eqnarray}
  \cZ^{(+)}_{\{(T_I,N_I)\}} &=& \prod_I \frac{T_I!}{N_I!(T_I - N_I)!} e^{-\beta(\epsilon_I-\mu) N_I}, \label{cZB_def} \\
  \cZ^{(-)}_{\{(G_I,N_I)\}} &=& \prod_I \frac{(G_I+N_I-1)!}{N_I!(G_I - 1)!} e^{-\beta(\epsilon_I-\mu) N_I}. \label{cZF_def}
\end{eqnarray}
\end{subequations}
In Eqs. (\ref{cZBF_part}) we made the usual simplifying assumption that all the chemical potentials in all the species are the same, i.e. $\mu_I\equiv\mu$ for all $I$. 
This assumption is justified by the fact that in the present
application all the particles are identical and, as we shall see below, a species represents a single particle energy interval.

The equilibrium populations are obtained by maximizing $\cZ^{(+)}_{\{(T_I,N_I)\}}$ and $\cZ^{(-)}_{\{(G_I,N_I)\}}$ with respect to $N_I$, taking into account the variation of the number of states in the species with the number of particles. We obtain the equations
\begin{equation}
  \ln\frac{1\mp n^{(\pm)}_K}{n^{(\pm)}_K} \pm \sum_I\alpha_{IK}\ln(1\mp n^{(\pm)}_I) - \beta(\epsilon_K-\mu) = 0 , \label{Eqs_ni_BF}
\end{equation}
where $n^{(-)}_I\equiv N_I/G_I$ and $n^{(+)}_I\equiv N_I/T_I$ \cite{EPL.90.10006.2010.Anghel}. 

A fermionic system may be transformed into a bosonic system if we define $G_I= T_I-N_I+1$ and $\alpha'_{IJ}=\alpha_{IJ} + \delta_{IJ}$, which leads to $n^{(-)}_I \equiv N_I/G_I \approx n^{(+)}_I/[1-n^{(+)}_I]$.

In the ideal system the total energy is equal to the sum of quasiparticle energies, which are independent of the populations. We define alternative quasiparticle energies--the FES quasiparticle energies--as
\begin{eqnarray}
  &&\epsilon_\rr \equiv t+V_{\rm ext}(\rr) + \int\limits_\Omega d^s\rr' \theta[t +V_{\rm ext}(\rr) -V_{\rm ext}(\rr')] \nonumber \\
  && \times\int\limits_0^{t+V_{\rm ext}(\rr)-V_{\rm ext}(\rr')} dt' \sigma(\rr',t') n(\rr',t') v(|\rr-\rr'|) . \label{til_eps_lr_FES}
\end{eqnarray}
and we can immediately check that
\begin{equation}
E=  \int_\Omega d^s\rr' \int_0^\infty  \epsilon_\rr  n^{(\pm)}[\rr,t(\epsilon_\rr)]\sigma(\rr,\epsilon_\rr) \, d\epsilon_\rr
\label{Etot}
\end{equation}
Equation (\ref{Etot}) shows that the total energy can be obtained as a simple sum of quasi-particles
energies, meaning that these latter can be viewed as an ideal gas. This
surprising result for an interacting system is well known in the context of FES. 
From Eq. (\ref{til_eps_lr_FES}) we obtain the DOS along the $\epsilon$ axis,
\begin{eqnarray}
  && \sigma[\rr,\epsilon_\rr(t)] = \sigma(\rr,t)\left|\frac{d\epsilon_\rr}{d t}\right|^{-1} = \sigma(\rr,t) \left|1+\int_\Omega d^s\rr' \right. \label{tilde_sigma_def} \\
  && \times \theta[V_{\rm ext}(\rr) -V_{\rm ext}(\rr')]v(|\rr-\rr'|) \sigma(\rr',t) n[\rr',t(\epsilon_\rr)]\bigg|^{-1}. \nonumber
\end{eqnarray}
Although $\epsilon_\rr$ depends explicitly on $n(\rr,t)$ (\ref{til_eps_lr_FES}), we can transfer this dependence to the statistical interaction through the FES parameters which leaves the quasiparticle gas an ideal gas, as explained, for example, in Ref. \cite{PhysScr.2012.014079.2012.Anghel}. The FES parameters are calculated similarly to Refs. \cite{PhysLettA.372.5745.2008.Anghel,PhysLettA.376.892.2012.Anghel}. To this aim in each such volume $\delta\rr$ the quasiparticle energy axis $\epsilon$ is split into elementary intervals $\delta\epsilon$ centered at $\epsilon$. Each $(s+1)$-dimensional elementary volume, $\delta\rr\times\delta\epsilon$, represents a FES species, as indicated in Fig. \ref{species}.
%
Between species with the same energy, $\epsilon\equiv\epsilon_3$ [arrow (a) in Fig. \ref{species}], but located in different volumes, $\delta{\rr_1}$ and $\delta{\rr_2}$, we have the FES parameters
\begin{subequations} \label{alphas_lr}
\begin{equation}
  \alpha_{\delta{\rr_1} \epsilon_3;\delta{\rr_2}\epsilon_3} = v(|\rr_1-\rr_2|) \sigma[\rr_1,t(\epsilon_3)] \delta{\rr_1} . \label{alpha1_lr}
\end{equation}
Between species with different energies, $\epsilon_2\ne\epsilon_3$ [arrow (b) in Fig. \ref{species}], we have the parameters
\begin{eqnarray}
  \alpha_{\delta{\rr_1}\epsilon_3;\delta{\rr_2}\epsilon_2} &=& \theta(\epsilon_3-\epsilon_2)v(|\rr_1-\rr_2|) \left.\frac{d\{\ln[\sigma(\rr_1,t)]\}}{dt}\right|_{t(\epsilon_3)} \nonumber \\
  && \times\sigma[\rr_1,t(\epsilon_1)] \delta{\rr_1} \delta t 
  \label{alpha2_lr}
\end{eqnarray}
%
Finally, if $\epsilon_2\ne\epsilon_1$ and $\epsilon_1$ is the lowest energy species in the volume $\delta{\rr_1}$, i.e. $t(\epsilon_1)=0$ [arrow (c) in Fig. \ref{species}], then we have
\begin{equation}
  \alpha_{\delta{\rr_1}\epsilon_1;\delta{\rr_2}\epsilon_2} = \theta(\epsilon_1 -\epsilon_2) v(|\rr_1-\rr_2|) \sigma(\rr_1,0) \delta{\rr_1} . \label{alpha3_lr}
\end{equation}
\end{subequations}
The presence of non-zero FES parameters is a proof that the system obeys FES. 

We now turn to show that the whole thermodynamics can be equivalently calculated in the TF and in the FES approaches.
Plugging the FES parameters (\ref{alphas_lr}) and the quasiparticle energies (\ref{til_eps_lr_FES}) into the FES equations (\ref{Eqs_ni_BF}) we obtain
\begin{eqnarray}
  && \beta(\mu-\epsilon_\rr) + \ln\frac{[1\mp n^{(\pm)}(\rr,\epsilon_\rr)]}{n^{(\pm)}(\rr,\epsilon_\rr)} \nonumber \\
  && = \mp \int_\Omega d^s\rr'\theta[\epsilon_\rr-V_{\rm ext}(\rr')] v(|\rr-\rr'|) \nonumber \\
  && \times \left\{\sigma[\rr',t(\epsilon_\rr)] \ln[1\mp n^{(\pm)}(\rr',\epsilon_\rr)] \right. \nonumber \\
  && \left. + \int_{\epsilon(\epsilon_\rr)}^\infty dt' \frac{\partial\sigma(\rr',t')}{\partial t'}\ln[1\mp n^{(\pm)}(\rr',t')] \right\} \nonumber \\
  && \mp \int_\Omega d^s\rr' \theta[V_{\rm ext}(\rr')-\epsilon_\rr] v(|\rr-\rr'|) \nonumber \\
  && \times \left\{\sigma(\rr',0)\ln[1\mp n^{(\pm)}(\rr',0)] \right. \nonumber \\
  && \left. + \int_{0}^\infty dt'\,\frac{\partial\sigma(\rr',t')}{\partial t'}\ln[1\mp n^{(\pm)}(\rr',t')] \right\} .
  \label{Eq_pop_lr}
\end{eqnarray}

Now we can check the equivalence between the FES and the TF descriptions of the system by substituting Eq. (\ref{npm_eq_L}) into Eq. (\ref{Eq_pop_lr}). By doing so we recover the relation between $\epsilon_\rr$ and {$\tilde\epsilon_\rr$},
%
\begin{eqnarray}
  \epsilon_\rr &\equiv& \tilde\epsilon_{\rr} - \int\limits_\Omega d^s\rr' \left\{ \theta[\epsilon_\rr-V_{\rm ext}(\rr')] \int_{t(\epsilon_\rr)}^\infty dt' \sigma(\rr',t') \right. \nonumber \\
  && \times n(\rr',t') v(|\rr-\rr'|) + \theta[V_{\rm ext}(\rr') -\epsilon_\rr] \nonumber \\
  && \times\int_0^\infty dt' \sigma(\rr',t') n(\rr',t') v(|\rr-\rr'|) \Bigg\} , \label{til_eps_consistent}
\end{eqnarray}
which is in accordance with the definitions (\ref{til_epsL}) and (\ref{til_eps_lr_FES}). 

Equations (\ref{til_epsL}) and (\ref{til_eps_lr_FES}), as well as Eqs. (\ref{Eq_pop_lr}) and (\ref{til_eps_consistent}), make sense only if the integrals in these expressions converge. This is a limitation of the mean-field formalism.

\section{Numerical example} \label{num_ex}

\begin{figure}[t]
\begin{center}
\includegraphics[width=8.5cm]{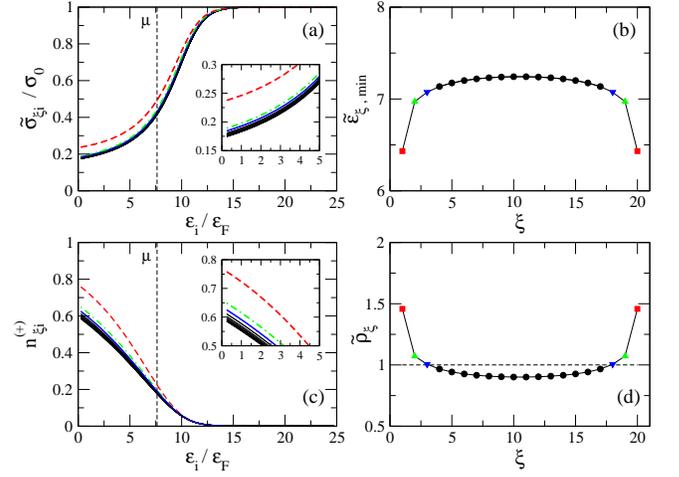}
\end{center}
\caption{(Color online)
Specific FES and TF quantities for a one dimensional system with repulsive Coulomb interactions
($N_\rr=20$, $N_{\epsilon}=50$, $\epsilon_{\rm max}=25$):
(a) the quasiparticle density of states in the FES description; 
(b) the position dependent energy shift $\tilde\epsilon_{\xi,{\rm min}}$ of Landau's quasiparticle energies;
(c) the FES populations; and
(d) the particle density, scaled with $\epsilon_F\sigma_0$--the dashed line represents
the uniform distribution of a similar non-interacting system.
In the (a) and (c) insets we plot a curve for each elementary segment $\delta\rr_\xi$, $\xi=1,\ldots,20$.
We use the same symbols (color) for symmetric segments, namely, square (red) for $1$ and $N_\rr$, 
up triangle (green) for $2$ and $N_\rr-1$, down triangle (blue) for $3$ and $N_\rr-2$, and circles (black) for the rest.
The chemical potential $\mu$ is marked by vertical dashed lines.
}
\label{sigma_pop}
\end{figure}

We illustrate our model and the relation between the FES and the TF descriptions
on a one dimensional system of fermions
with repulsive Coulomb interactions, $V(|\rr-\rr'|) = 1/|\rr-\rr'|$, in the absence of external fields. The length of the system, $\Omega$, is discretized as in Ref. \cite{JPhysConfSer.410.012120.2013.Nemnes} into $N_\rr$ equal elementary segments, $\delta\rr_\xi=1$, where $\xi=1,\ldots,N_\rr$. In each such elementary ``volume'' the DOS is taken to be constant,  {$\delta\rr_\xi\sigma(\rr_\xi,t)\equiv\sigma_0$}, and on the $\epsilon$ axis we define $N_{\epsilon}$ equal consecutive segments between 0 and $\epsilon_{\rm max}$, $\delta\epsilon_i\equiv \epsilon_{\rm max}/N_{\epsilon}$, where $i=1,\ldots,N_{\epsilon}$ -- we choose $\epsilon_{\rm max}$ such that $n^{(+)}(\rr,\epsilon_{\rm max})\ll1$ for any $\rr$. 
In this way we obtain $N_\rr \times N_{\epsilon}$ species of particles, $\delta\rr_\xi\times\delta\epsilon_i$, identified also by a double index, $(\xi,i)$ \cite{JPhysConfSer.410.012120.2013.Nemnes}. To avoid the singularity at the origin of the interaction potential, we consider a cut-off distance of $\delta\rr_\xi/2$.

The total number of particles in the system is $N=N_\rr \epsilon_F \sigma_0$, where $\epsilon_F$ is the Fermi energy in the noninteracting system. We set the energy scale of the system by fixing $\epsilon_F=1$ and $k_BT=1/\beta=1$.

Figure \ref{sigma_pop}(a) shows the quasiparticle density of states in the FES description, $\sigma_{\xi i} \equiv\delta\rr_\xi \sigma(\rr_\xi,\epsilon_i)$ (Eq. \ref{tilde_sigma_def}). The density of Landau's quasiparticle states may be calculated also as $\tilde\sigma[\rr,\tilde\epsilon_\rr(t)] = \sigma(\rr,t)\left|d\tilde\epsilon_\rr/d t\right|^{-1}$ and we obtain $\tilde\sigma_{\xi i}(\tilde\epsilon) = \theta(\tilde\epsilon - \tilde\epsilon_{\xi,{\rm min}})\sigma_0$, which is different from zero only for $\tilde\epsilon \ge\tilde\epsilon_{\xi,{\rm min}}$. The minimum value of $\tilde\epsilon_\xi$ (\ref{til_epsL}) is $\tilde\epsilon_{\xi,{\rm min}}\equiv \tilde\epsilon_{\rr_\xi}(t=0) =\tilde\epsilon_{\rr_\xi}(t)-t$. The position dependent $\tilde\epsilon_{\xi,{\rm min}}$ is plotted in Fig. \ref{sigma_pop} (b). 

Figure \ref{sigma_pop}(c) shows the populations, $n^{(+)}_{\xi i} \equiv n^{(+)}(\rr_\xi,\epsilon_i)$ (\ref{Eq_pop_lr}). To obtain Landau's populations (\ref{npm_eq_L}), we simply shift the quasiparticle energy from $\epsilon_i$ (\ref{til_eps_lr_FES}) to $\tilde\epsilon_{\xi i}$ (\ref{til_epsL}) and $n^{(+)}[\rr_\xi,\tilde\epsilon(\epsilon_{\xi i})]$ becomes a Fermi distribution in $\tilde\epsilon_{\xi i}$ for any $\xi$ and with the same $\mu$ as for the FES distribution. For example one may substitute $\tilde\epsilon_{\xi,{\rm min}}$ into Eq. (\ref{npm_eq_L}) and obtain the same population for the species with the lowest energy, $n^{(+)}_{\xi,{i=0}}$, presented in the FES description in Fig.\ \ref{sigma_pop}(c).

The particle density, $\tilde\rho_\xi = \sum_i \tilde\sigma_{\xi i} n_{\xi i} \delta\tilde\epsilon_i$, is represented in Fig.~\ref{sigma_pop}(d). Due to the symmetry of the problem, we have pair-wise identical populations, $n_{\xi, i}=n_{N_\rr-\xi+1, i}$ and $\tilde\rho_\xi=\tilde\rho_{N_\rr-\xi+1}$, for any $\xi$ and $i$. The largest deviations in the particle density occur for the extremal species, placed at both ends of the one-dimensional (1D) box as an effect of repulsive interactions.

We observe from Eqs. (\ref{til_epsL}) and (\ref{til_eps_lr_FES}) and from Fig.~\ref{sigma_pop} that because $V_{\rm ext}(\rr)=0$, the range of $\epsilon_{\rr_\xi}$ is $[0,\infty)$ in any elementary volume $\xi$, whereas the range of $\tilde\epsilon_{\rr_\xi}$ is $[\tilde\epsilon_{\xi,{\rm min}},\infty)$, with $\tilde\epsilon_{\xi,{\rm min}}>0$. 

\section{Analytical examples} \label{O_ex}

\paragraph{Calogero-Sutherland model in a one-dimensional harmonic trap.}

In Ref. \cite{PhysRevLett.73.3331.1994.Murthy} Murthy and Shankar analyzed the Calogero-Sutherland model (CSM), i.e. a 1D system of fermions in a harmonic potential of frequency $\omega$, with inverse square law particle-particle interaction potential, $v(r)\propto r^{-2}$,
\begin{equation}
  H = \sum_{i=1}^N \left( -\frac{1}{2}\frac{\partial^2}{\partial x_i}\right) + \frac{1}{2} \sum_{i<j=1}^N \frac{g(g-1)}{(x_i-x_j)^2} , \label{CSM_H}
\end{equation}
where $N$ is the fixed total number of particles and we take, like in Ref. \cite{PhysRevLett.73.3331.1994.Murthy}, $\hbar=m=1$, $m$ being the particle mass. 
Such a system is not solvable in the TF approximation, but its spectrum is exactly known \cite{PhysRevLett.73.3331.1994.Murthy,Sutherland:book}:
\begin{equation}
  E = \sum_{k=0}^\infty t_k n_k - \omega(1-g)\frac{N(N-1)}{2} , \label{CSM_E}
\end{equation}
where $t_k = k\omega$, $k$ is an integer, $n_k$ is the occupation number, and $N=\sum_{k=0}^\infty n_k$. The energy (\ref{CSM_E}) is of mean-field type (\ref{E_FLT}) and from this point the formalism of Section \ref{FES_section} may be applied straightforwardly, with $v(r) = -\omega(1-g)$, independent of $r$. The quasiparticle energies are \cite{PhysRevLett.73.3331.1994.Murthy}
\begin{equation}
  \epsilon_k = t_k - \omega (1-g) \sum_{l(<k)} n_l \label{qp_en_CSM1}
\end{equation}
and the species are small intervals $\delta\epsilon$ centered at $\epsilon$, along the quasiparticle energy axis. In this way, from (\ref{alpha1_lr}) and observing that $\omega=\sigma^{-1}$, we get the ``diagonal'' FES parameters,
\begin{equation}
  \alpha_{\epsilon;\epsilon} = g-1 . \label{alpha_CSM}
\end{equation}
From (\ref{alpha2_lr}) we get $\alpha_{\epsilon;\epsilon'} = 0$ for any $\epsilon\ne\epsilon'$, since the DOS is constant in this case.

Finally, Eq. (\ref{alpha3_lr}) is not applicable to this system since we work with single-particle states extended over the whole system and not in the TF approximation. In conclusion we can write, in general, that $\alpha_{\epsilon;\epsilon'} = (g-1)\delta_{\epsilon,\epsilon'}$. 

This result is identical with that of Murthy and Shankar, considering that we calculate $\alpha_{\epsilon;\epsilon'}$ in the fermionic picture, whereas $\alpha'_{\epsilon;\epsilon'} =g\delta_{\epsilon;\epsilon'}$ of Ref. \cite{PhysRevLett.73.3331.1994.Murthy} was calculated in the bosonic picture. The two $\alpha$'s should satisfy the relation $\alpha'_{\epsilon;\epsilon'} = \alpha_{\epsilon;\epsilon'} + \delta_{\epsilon;\epsilon'}$ (see Section \ref{FES_section}), which is correct.

\paragraph{Calogero-Sutherland model on a ring.}

FES may be applied not only in the energy space, but also in the (quasi)momentum space \cite{NewDevIntSys.1995.Bernard,PhysRevB.56.4422.1997.Sutherland,PhysRevLett.85.2781.2000.Iguchi}. Following Ref. \cite{PhysRevB.56.4422.1997.Sutherland} (and keeping $\hbar=c=m=1$) for a CSM system of $N$ particles on a ring of length $L$, the equation for the asymptotic momentum $k$ is
\begin{equation}
  Lk - \sum_{k'}\phi(k-k') = 2\pi I(k) \equiv L k_0. \label{CSM_k1}
\end{equation}
The sum is taken over all the particles in the system, $I(k)$ is an integer, and $\phi(k-k')$ is the phase shift due to the particle-particle interaction. The total number of particles, the momentum, and the energy of the system are
\begin{equation}
  N = \sum_k 1,\ P = \sum_k k,\ {\rm and}\ E = \sum_k \frac{k^2}{2}, \label{CSM_kNPE}
\end{equation}
respectively. Since $I$ takes integer values, Eq. (\ref{CSM_k1}) leads to a density of states along the $k$ axis \cite{EPL.90.10006.2010.Anghel},
\begin{equation}
  \sigma(k) = \frac{L}{2\pi}\left\{1-\frac{1}{L}\int\phi'(k-k')\sigma(k')n(k')\,dk'\right\} . \label{defsigmak1} 
\end{equation}
The species are defined as intervals $\delta k$, centered at $k$, along the momentum axis, and the FES parameters are 
\begin{equation}
  \alpha_{k;k'} = \frac{1}{2\pi}\phi'(k-k')\delta k , \label{tildealpha}
\end{equation}
where $\phi'(k)=d\phi(k)/dk$, and $n(k)$ is the occupation of the state with asymptotic momentum $k$. 
If the interaction is like in the previous example, $v(|x-x'|)=g(g-1)/(x-x')^2$, then $\phi(k)=\pi(g-1){\rm sgn}(k)$, with ${\rm sgn}(k)$ being the sign of $k$, and we obtain again $\alpha_{k;k'}=(g-1)\delta_{k;k'}$ \cite{NewDevIntSys.1995.Bernard,PhysRevB.56.4422.1997.Sutherland,EPL.90.10006.2010.Anghel}.


The problem may be transferred from the momentum space to the quasiparticle energy space. The total energy of the system is $E$ (\ref{CSM_kNPE}) and the quasiparticle energy is
\begin{equation}
  \epsilon = \frac{k^2}{2} . \label{qp_en_CSM2}
\end{equation}
The species $\delta k$ along the $k$ axis are mapped into species $\delta\epsilon$ along the $\epsilon$ axis. To each species $\delta\epsilon$, there corresponds two species, $\delta k$ and $\delta k'$, symmetric with respect to the origin on the $k$ axis -- that is, if $\delta k$ is centered at $k$, then $\delta k'$ is centered at $-k$ and $\epsilon=k^2/2$. Therefore every two symmetric species, $\delta k$ and $\delta k'$ are combined into one energy species, $\delta\epsilon$. 
If the dimensions of the species on the $k$ axis are $G_{\delta k}$ and $G_{\delta k'}$, respectively, then the dimension of the species $\delta\epsilon$ is $G_{\delta\epsilon} = G_{\delta k} + G_{\delta k'}$. A similar relation holds for the particle numbers: $N_{\delta\epsilon} = N_{\delta k} + N_{\delta k'}$. The FES parameters in the $\epsilon$ space are obtained by applying the rules of Ref. \cite{EPL.87.60009.2009.Anghel}. If we calculate the $\alpha_{\epsilon_1,\epsilon_2}$, which connects the species 
$\delta\epsilon_2$ to the species $\delta\epsilon_1$, then the following relations have to be satisfied:
\begin{equation}
  \alpha_{\epsilon_1,\epsilon_2} = \alpha_{k_1,k_2} +  \alpha_{-k_1,k_2} = \alpha_{k_1,-k_2} +  \alpha_{-k_1,-k_2}, \label{FES_rel1}
\end{equation}
where $\delta\epsilon_1$ corresponds to the intervals $\delta k_1$ and $\delta k_1'$ on the $k$ axis, whereas $\delta\epsilon_2$ corresponds to the intervals $\delta k_2$ and $\delta k_2'$.  If $\alpha_{k;k'}=(g-1)\delta_{k;k'}$, then
\begin{equation}
  \alpha_{\epsilon_1,\epsilon_2} = (g-1)\delta_{k_1;k_2} + (g-1)\delta_{-k_1;k_2} \equiv (g-1) \delta_{\epsilon_1;\epsilon_2} , \label{FES_rel2}
\end{equation}
like in the case of the CSM in a harmonic trap. 

\section{Conclusions} \label{Conclusions}

In conclusion we have formulated an approach by which a system of {quantum} particles with general particle-particle interaction $V(|\rr-\rr'|)$ in an $s$-dimensional space and external potential $V_{\rm ext}(\rr)$ is described {in the quasiclassical limit} as an ideal gas of FES. We have given the equations for the calculation of the FES parameters and equilibrium populations.

The FES approach has been compared with the TF formalism and we have shown that although there are differences in the definitions of certain quantities like the quasiparticle energies, the physical results are the same.
The main difference between the two formalisms is that in the FES approach the quasiparticle energies are independent of the populations of other quasiparticle states and therefore the FES gas is ``ideal'', with the total energy of the gas being equal to the sum of the quasiparticle energies, whereas in the TF approach the quasiparticles are interacting and the energy of the quasiparticle gas is not equal to the energy of the system.

We have exemplified our procedure on a one-dimensional system of fermions with repulsive Coulomb interaction for which we calculated the main microscopic parameters, like the quasiparticle energies, quasiparticle density of states, and energy levels populations. For each of these quantities we discussed the similarities and differences between the FES and the TF approaches. 
We also applied our procedure on the one-dimensional CSM which is well studied in the literature \cite{PhysRevLett.73.3331.1994.Murthy,NewDevIntSys.1995.Bernard,PhysRevB.56.4422.1997.Sutherland,PhysRevLett.74.3912.1995.Sen} and proved that the results are consistent.

One practical consequence that appears from our calculations is that the solution of the FES integral equations may eventually be calculated easier by solving self-consistently the TF equations for population and quasiparticle energies, (\ref{npm_eq_L}) and (\ref{til_epsL}).

Another consequence is that while in the TF formulation the quasiparticle energies may form an energy gap at the lowest end of the spectrum due to the particle-particle interaction, in the FES description such an energy gap does not exist.

By establishing the equivalence between the self-consistent mean-field theory and the FES approach we show that in general a quasi-classical interacting system can be mapped onto an ideal FES system.

\section{Acknowledgements}

The work was supported by the Romanian National Authority for Scientific Research CNCS-UEFISCDI Projects No. PN-II-ID-PCE-2011-3-0960 and No. PN09370102/2009. The travel support from the Romania-JINR Dubna Collaboration Project Titeica-Markov is gratefully acknowledged. 


\begin{thebibliography}{10}

\bibitem{NuclPhysB92.445.1975.Chaichian}
M.~Chaichian, R.~Hagedorn, and M.~Hayashi.
\newblock {Nucl. Phys. B} {\bf 92}, 445 (1975).

\bibitem{PhysRevLett.72.600.1994.Veigy}
A.~Dasni{\`e}res~de Veigy and S.~Ouvry.
\newblock {Phys. Rev. Lett.} {\bf 72}, 600 (1994).

\bibitem{PhysRevLett.73.922.1994.Wu}
Y.-S. Wu.
\newblock {Phys. Rev. Lett.} {\bf 73}, 922 (1994).

\bibitem{PhysRevLett.73.2150.1994.Isakov}
S.~B. Isakov.
\newblock {Phys. Rev. Lett.} {\bf 73(16)}, 2150 (1994).

\bibitem{FES_intro2013.Murthy}
M.~V.~N. Murthy and R.~Shankar.
\newblock Exclusion statistics: From pauli to haldane.
\newblock Report of The Institute of Mathematical Sciences, Chennai, India,
  MatSciRep:120, 2013.

\bibitem{PhysRevLett.67.937.1991.Haldane}
F.~D.~M. Haldane.
\newblock {Phys. Rev. Lett.} {\bf 67}, 937 (1991).

\bibitem{NewDevIntSys.1995.Bernard}
D.~Bernard and Y.~S. Wu.
\newblock In M.~L. Ge and Y.~S. Wu, editors, {\it New Developments on
  Integrable Systems and Long-Ranged Interaction Models}, page~10. World
  Scientific, Singapore, 1995.
\newblock cond-mat/9404025.

\bibitem{PhysRevE.75.61120.2007.Potter}
G.~G. Potter, G~M{\"u}ller, and M~Karbach.
\newblock {Phys. Rev. E} {\bf 75}, 61120 (2007).

\bibitem{PhysRevE.76.61112.2007.Potter}
G.~G. Potter, G~M{\"u}ller, and M~Karbach.
\newblock {Phys. Rev. E} {\bf 76}, 61112 (2007).

\bibitem{PhysRevE.84.021136.2011.Liu}
Dan Liu, Ping Lu, Gerhard M\"{u}ller, and Michael Karbach.
\newblock {Phys. Rev. E} {\bf 84}, 021136 (2011).

\bibitem{PhysRevE.85.011144.2012.Liu}
Dan Liu, Jared Vanasse, Gerhard M\"{u}ller, and Michael Karbach.
\newblock {Phys. Rev. E} {\bf 85}, 011144 (2012).

\bibitem{JStatMech.P04018.2013.Gundlach}
N.~Gundlach, M.~Karbach, D.~Liu, and G.~M\"{u}ller.
\newblock {J. Stat. Mech.: Theory and Experiment} {\bf 2013}, P04018 (2013).

\bibitem{PhysRevB.56.4422.1997.Sutherland}
B.~Sutherland.
\newblock {Phys. Rev. B} {\bf 56}, 4422 (1997).

\bibitem{PhysRevLett.85.2781.2000.Iguchi}
K.~Iguchi and B.~Sutherland.
\newblock {Phys. Rev. Lett.} {\bf 85}, 2781 (2000).

\bibitem{JPhysA.40.F1013.2007.Anghel}
D.~V. Anghel.
\newblock {J. Phys. A: Math. Theor.} {\bf 40}, F1013 (2007).

\bibitem{PhysLettA.372.5745.2008.Anghel}
D.~V. Anghel.
\newblock {Phys. Lett. A} {\bf 372}, 5745 (2008).

\bibitem{PhysLettA.376.892.2012.Anghel}
D.~V. Anghel.
\newblock {Phys. Lett. A} {\bf 376}, 892 (2012).

\bibitem{EPL.90.10006.2010.Anghel}
D.~V. Anghel.
\newblock {EPL} {\bf 90}, 10006 (2010).
\newblock arXiv:0909.0030.

\bibitem{PhysRevLett.74.3912.1995.Sen}
D.~Sen and R.~K. Bhaduri.
\newblock {Phys. Rev. Lett.} {\bf 74}, 3912 (1995).

\bibitem{JPhysB33.3895.2000.Bhaduri}
R.~K. Bhaduri, S.~M. Reimann, S.~Viefers, A.~G. Choudhury, and M.~K.
  Srivastava.
\newblock {J. Phys. B} {\bf 33}, 3895 (2000).

\bibitem{PhysRevLett.86.2930.2001.Hansson}
T.~H. Hansson, J.~M. Leinaas, and S.~Viefers.
\newblock {Phys. Rev. Lett.} {\bf 86}, 2930 (2001).

\bibitem{PhysRevE.82.031137.2010.Mirza}
B.~Mirza and H.~Mohammadzadeh.
\newblock Thermodynamic geometry of fractional statistics.
\newblock {Phys. Rev. E} {\bf 82}, 031137 (2010).

\bibitem{PhysRevE.83.021111.2011.Qin}
F.~Qin and J.-S. Chen.
\newblock {Phys. Rev. E} {\bf 83}, 021111 (2011).

\bibitem{PhysRevE.76.061123.2007.Pellegrino}
F.~M.~D. Pellegrino, G.~G.~N. Angilella, N.~H. March, and R.~Pucci.
\newblock {Phys. Rev. E} {\bf 76}, 061123 (2007).

\bibitem{JPhysA.26.4017.1993.Chaichian}
M.~Chaichian, R.~G. Felipe, and C.~Montonen.
\newblock {J. Phys. A: Math. Gen.} {\bf 26}, 4017 (1993).

\bibitem{JStatMech.P09011.2010.Nemnes}
G.~A. Nemnes and D.~V. Anghel.
\newblock {J. Stat. Mech.} {\bf 2010}, P09011 (2010).

\bibitem{JStatMech.2009.P02021.2009.Guastella}
I.~Guastella, L.~Bellomonte, and R.~M. Sperandeo-Mineo.
\newblock {J. Stat. Mech.} {\bf 2009}, P02021 (2009).

\bibitem{JPhysB42.235302.2009.Bhaduri}
R.~K. Bhaduri, M.~V.~N. Murthy, and M.~K. Srivastava.
\newblock {J. Phys. B: At. Mol. Opt. Phys.} {\bf 42}, 235302 (2009).

\bibitem{JPhysB.43.055302.2010.Qin}
F.~Qin and J.~Chen.
\newblock {J. Phys. B: Atomic, Molecular and Optical Physics} {\bf 43}, 055302 (2010).

\bibitem{JPhysA.45.315302.2012.vanZyl}
B.~P. van Zyl.
\newblock {J. Phys. A: Math. Theor.} {\bf 45}, 315302 (2012).

\bibitem{JPhysA.46.045001.2013.MacDonald}
Z.~MacDonald and B.~P. van Zyl.
\newblock {J. Phys. A: Math. Theor.} {\bf 46}, 045001 (2013).

\bibitem{EPL.87.60009.2009.Anghel}
D.~V. Anghel.
\newblock {EPL} {\bf 87}, 60009 (2009).
\newblock arXiv:0906.4836.

\bibitem{PhysRevLett.104.198901.2010.Anghel}
D.~V. Anghel.
\newblock {Phys. Rev. Lett.} {\bf 104}, 198901 (2010).

\bibitem{PhysRevLett.104.198902.2010.Wu}
Yong-Shi Wu.
\newblock {Phys. Rev. Lett.} {\bf 104}, 198902 (2010).

\bibitem{JPhysConfSer.410.012121.2013.Anghel}
D.~V. Anghel.
\newblock {J. Phys. Conf. Ser.} {\bf 410}, 012121 (2013).

\bibitem{PhysScr.2012.014079.2012.Anghel}
D.~V. Anghel.
\newblock {Physica Scripta} {\bf 2012}, 014079 (2012).

\bibitem{PhysRevLett.73.3331.1994.Murthy}
M.~V.~N. Murthy and R.~Shankar.
\newblock {Phys. Rev. Lett.} {\bf 73}, 3331 (1994).

\bibitem{JPhysConfSer.410.012120.2013.Nemnes}
G.~A. Nemnes and D.~V. Anghel.
\newblock {J. Phys.: Conf. Ser.} {\bf 410}, 012120 (2013).

\bibitem{Sutherland:book}
{\it Beautiful Models}.
\newblock World Scientific Publishing Co. Pte. Ltd., 2004.

\end{thebibliography}

\end{document}